\documentclass[aps,twocolumn,floats,prl,nofootinbib]{revtex4}
\usepackage{graphics,graphicx,epsfig}
\usepackage{amssymb,color}
\usepackage{amsmath}
\usepackage{epsf,epstopdf,wrapfig}

\begin{document}

\title{Information costs in the control of protein synthesis}
\author{Rebecca J. Rousseau\:$^{1,2}$ and William Bialek\:$^{1,3}$}
\date{\today}
\affiliation{$^1$Joseph Henry Laboratories of Physics, and Lewis--Sigler Institute for Integrative Genomics, Princeton University, Princeton NJ 08544\\
$^2$Department of Physics, California Institute of Technology, Pasadena CA 91125\\
$^{3}$Initiative for the Theoretical Sciences, The Graduate Center, City University of New York, 365 Fifth Ave, New York, NY 10016}

\date{\today}

\begin{abstract}
Efficient protein synthesis depends on the availability of charged tRNA molecules. With 61 different codons, shifting the balance among the tRNA abundances can lead to large changes in the protein synthesis rate. Previous theoretical work has asked about the optimization of these abundances, and there is some evidence that regulatory mechanisms bring cells close to this optimum, on average. We formulate the tradeoff between the precision of control and the efficiency of synthesis, asking for the maximum entropy distribution of tRNA abundances consistent with a desired mean rate of protein synthesis. Our analysis, using data from {\em E coli}, indicates that reasonable synthesis rates are consistent only with rather low entropies, so that the cell's regulatory mechanisms must encode a large amount of information about the ``correct" tRNA abundances.
\end{abstract}

\maketitle

In order to function efficiently, living cells need to take control over their internal chemistry. This involves adjusting the concentration of relevant molecules in relation to some goal, and requires transmitting information about the goal through the regulatory elements that control these concentrations.  But in many biological regulatory mechanisms, information in turn is represented by the concentrations of other molecules, and these concentrations often are quite low, leading to physical limits on information transmission \cite{bialek_12}. Here we explore the tradeoff between information and efficiency in the context of protein synthesis.

Protein synthesis requires transfer RNA (tRNA) molecules to arrive at the ribosome and dock with their complementary codons along the messenger RNA (mRNA).  It was realized some time ago that maximizing the rate of protein synthesis requires matching tRNA abundances to codon usage, and there is some evidence that this happens, at least on average \cite{xia_88,solomovici+al_97}.  A similar idea has been applied to bacterial metabolism as a whole, where the fluxes of individual biochemical steps can be tuned to maximize the conversion nutrients into biomass \cite{flux_opt}.  But these discussions of optimization assume that the cell can fix molecular abundances with infinite precision.  In a series of papers, De Martino and colleagues have constructed maximum entropy models for the distribution of metabolic fluxes that are consistent with a given mean rate of conversion into biomass \cite{demartino+al_16}.  As with networks of neurons \cite{neurons_maxent}, flocks of birds \cite{birds_maxent}, families of protein sequences \cite{proteins_maxent}, and more, these maximum entropy models constrained by low order moments provide surprisingly accurate, quantitative descriptions of emergent behavior in the metabolic network \cite{demartino+al_18}.

Here we use the maximum entropy construction to analyze the information required for efficient protein synthesis.  Concretely, we want to find the maximum entropy distribution of tRNA abundances that is consistent with a given mean rate of protein synthesis.  Our interest is not (immediately) in the distribution itself, but rather in the entropy.  We recall that reductions in entropy correspond to a gain in information, and by definition the maximum entropy model gives the smallest entropy reduction needed to satisfy the constraints \cite{jaynes_57}.  Thus our goal is to find the minimum amount of information required to specify the range of tRNA abundances that allow for protein synthesis at a given average rate.  This discussion is in the same spirit as classical analyses of the tradeoffs among growth rate, accuracy of information transmission, and metabolic costs \cite{EhrenbergKurland1984}.

The time required for protein synthesis must be longer than the time for charged tRNA molecules to arrive at the ribosome, and for each codon this time is inversely proportional to the tRNA concentration; it seems likely that tRNA availability in fact is the rate--limiting factor in translation elongation \cite{limiting}.  We can choose units where the average of this time, normalized per codon, is
\begin{equation}
T = \sum_{\rm i} {{f_{\rm i}}\over {t_i}} ,
\end{equation}
where $\rm i$ indexes the $K$ different codons, $f_{\rm i}$ is the fractional abundance of codon $\rm i$ in the synthesized proteins, and $t_{\rm i}$ is the abundance of the corresponding tRNA. Here we imagine that each codon has its own dedicated tRNA molecule, and return to the more realistic case below. Again we can choose units so that the mean total abundance of tRNA is normalized,
\begin{equation}
\sum_{\rm i} \langle t_{\rm i} \rangle = 1 .
\label{mean-t}
\end{equation}
In general, if we fix the mean of several functions $f_\mu (\{t_{\rm i}\})$ then the maximum entropy distribution is
\begin{equation}
P(\{t_{\rm i}\})) = {1\over {Z(\{g_\mu\})}} \exp\left[-\sum_\mu g_\mu f_\mu (\{t_{\rm i}\}) \right],
\end{equation}
where the Lagrange multipliers $\{g_\mu\}$ must be set so that the expectation values $\langle f_\mu (\{t_{\rm i}\})\rangle$ satisfy the constraints we have set \cite{jaynes_57}.  In our case, then, 
\begin{equation}
P( \{ t_{\rm i} \}) = {1\over {Z(\lambda, \mu)}} 
\exp\left[ - \lambda \sum_{\rm i} {{f_{\rm i}}\over {t_i}} - \mu \sum_{\rm i} t_{\rm i} 
\right]  ,
\label{P}
\end{equation}
where $\lambda$ fixes the mean synthesis time $\langle T\rangle$ and $\mu$ fixes the mean total tRNA abundance.

We have the usual ``thermodynamic'' identities,
\begin{eqnarray}
\langle T \rangle &=& - {{\partial \ln Z(\lambda, \mu )}\over{\partial \lambda}} ,\\
\sum_{\rm i} \langle t_{\rm i} \rangle &=& - {{\partial \ln Z(\lambda, \mu )}\over{\partial \mu}} =1,
\label{meant_con}
\end{eqnarray}
and the entropy of the distribution is
\begin{equation}
S = \ln Z(\lambda, \mu ) + \lambda \langle T \rangle + \mu .
\end{equation}
Because the constraints we have imposed do not require correlations among the different tRNA abundances, we can write the partition function exactly as a product,
\begin{eqnarray}
Z(\lambda , \mu) &=& \prod_{\rm i} \int dt\, e^{-\phi_{\rm i}(t)}\\
\phi_{\rm i} (t ) &=& \lambda f_{\rm i}/t  + \mu t .  
\label{phi_def}
\end{eqnarray}
We notice that 
\begin{equation}
\int dt\, e^{-\phi_{\rm i}(t)} = 2 \sqrt{\lambda f_{\rm i}/\mu} K_1(2\sqrt{\lambda\mu f_{\rm i}}),
\label{bessel}
\end{equation}
where $K_1(z)$ is the modified Bessel function of the second kind \cite{bessel_ref}.

We are especially interested in constraints that are strong enough to drive $T$ close to its minimum value.  In this limit, which is found at large $\lambda$, the distribution will be well approximated as a Gaussian around the minimum of each $\phi_{\rm i} $, 
\begin{equation}
P( \{ t_{\rm i} \})  = \prod_{\rm i} {1\over\sqrt{2\pi \sigma_{\rm i}^2}} \exp\left[ - {{(t_{\rm i} - t_{\rm i}^*)^2}\over{2\sigma_{\rm i}^2}}\right],
\end{equation}
where $t_{\rm i}^*$ is value of $t_{\rm i}$ that minimizes $\phi_{\rm i} (t)$, and 
\begin{equation}
{1\over{\sigma_{\rm i}^2}} = {{\partial^2 \phi_{\rm i} (t)}\over{\partial t^2}}{\bigg |}_{t = t_{\rm i}^*} .
\end{equation}
The entropy of this multidimensional Gaussian is then
\begin{equation}
S = {1\over 2}\sum_{\rm i} \log_2 (2\pi e \sigma_{\rm i}^2) \,{\rm bits} .
\label{entropy1}
\end{equation}

From Eq (\ref{phi_def}) we find that $t_{\rm i}^* = \sqrt{\lambda f_{\rm i}/\mu}$, and to obey Eq (\ref{mean-t}) we then must have
\begin{eqnarray}
\mu &=& \lambda \left( \sum_{\rm i} f_{\rm i}^{1/2} \right)^2 \\
t_{\rm i}^* &=& {{f_{\rm i}^{1/2}}\over{\sum_{\rm j} f_{\rm j}^{1/2}}} .
\end{eqnarray}
This scaling of the optimal tRNA abundances with the square--root of codon usage is familiar from previous work \cite{xia_88}.  At these optimal abundances we find the minimum synthesis time,
\begin{equation}
T_{\rm min} = \sum_{\rm i} {{f_{\rm i}}\over {t_i^*}} = \left( \sum_{\rm j} f_{\rm j}^{1/2}\right)^{2} .
\end{equation}

Similarly we have
\begin{equation}
 \sigma_{\rm i}^2 = {{t^3}\over{2\lambda f_{\rm i}}}{\bigg |}_{t = t_{\rm i}^*} 
= {1\over{2\lambda T_{\rm min}}} {{f_{\rm i}^{1/2}}\over{\sum_{\rm j} f_{\rm j}^{1/2}}} .
\end{equation}
If we compute the average synthesis time in this Gaussian distribution we find, to leading order in the variances $\sigma_{\rm i}^2$,
\begin{eqnarray}
\langle T \rangle &=& T_{\rm min} +  \sum_{\rm i} {{f_{\rm i}}\over{(t_{\rm i}^*)^3}} \sigma_{\rm i}^2 + \cdots \\
&=&  T_{\rm min} + {K\over{2\lambda}} .
\end{eqnarray}
Thus we have
\begin{equation}
{1\over {2\lambda T_{\rm min}} }  = {1\over K} {{\langle T \rangle - T_{\rm min}}\over{T_{\rm min}}} .
\end{equation}

Substituting into the entropy from Eq (\ref{entropy1}), we obtain
\begin{equation}
S = {1\over 2} \sum_{\rm i}\log_2\left[ {{2\pi e}\over K}   {{f_{\rm i}^{1/2}}\over{\sum_{\rm j} f_{\rm j}^{1/2}}}   {{\langle T \rangle - T_{\rm min}}\over{T_{\rm min}}} \right] .
\end{equation}
This illustrates the basic tradeoff between synthesis time and entropy:  if the cell wants to drive $\langle T \rangle \rightarrow T_{\rm min}$, then the entropy of the distribution of tRNA abundances must become smaller and smaller, corresponding to tighter control.  To set scale of this effect we compare with the entropy that is possible when we constrain the mean tRNA abundances but place no constraint on the synthesis times.  This corresponds to the distribution in Eq (\ref{P}) with $\lambda = 0$ and $\mu = K$; for this exponential distribution we can evaluate the entropy exactly,
\begin{equation}
S_0 = K \log_2 (e/K).
\end{equation}
Finally, the difference between $S_0$ and $S$ is the information required to specify the tRNA concentrations, 
\begin{equation}
I = {1\over 2} \sum_{\rm i}\log_2\left[ {e\over {2\pi K f_{\rm i}^{1/2}}}   \left(\sum_{\rm j} f_{\rm j}^{1/2}\right)    {{T_{\rm min}}\over{\langle T \rangle - T_{\rm min}}} \right] \,{\rm bits} .
\label{Igauss}
\end{equation}

Roughly speaking, holding the system within some desired distance of optimal synthesis rates requires keeping the variance of tRNA abundances small, proportional to the distance from the optimum.  But entropies are related to (half) the log of the variance, and this is true for each of the codons, giving us the form of Eq (\ref{Igauss}).  In more detail, we see that if the $f_{\rm i}$ are uniform, then $\sum_{\rm j} f_{\rm j}^{1/2}$ cancels $K f_{\rm i}^{1/2}$, and the log depends {\em only} on the distance from the optimum; the number of codons then simply scales the overall information.

\begin{figure}
\centerline{\includegraphics[width = 0.9\linewidth]{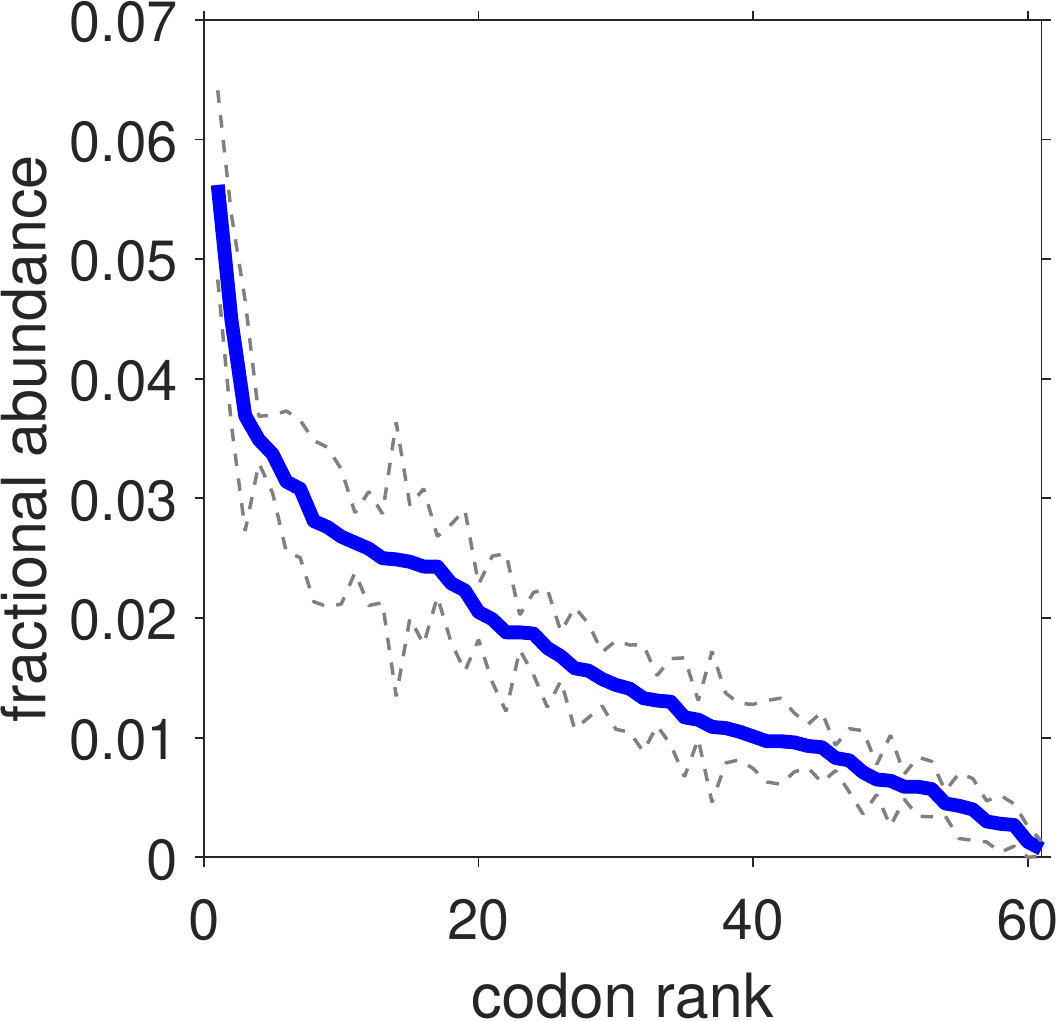}}
\caption{Normalized codon frequencies, $\{f_{\rm i}\}$, inferred from measurements of protein abundances in {\em E coli} under standard glucose conditions \cite{caglar+al_17}.  We show the mean (solid) $\pm$ one standard deviation (dashed) across the three separate experiments at nine timepoints during growth. 	\label{fi_fig}}
\end{figure}

In order to apply these ideas to real cells, we need to know the codon abundances $\{f_{\rm i}\}$.  It is easy to read the codons as they occur in the genome, but what matters here is the frequency with which they are used in making proteins.  Recent measurements on the bacterium {\em E coli}  survey the relative concentrations of all the expressed proteins under a variety of growth conditions \cite{caglar+al_17}, and we can use these results to estimate the codon abundances, with results shown in Fig \ref{fi_fig}.    We see that the $f_{\rm i}$ are far from uniform, varying over nearly two orders of magnitude, as known from earlier work \cite{Sharp1993,Xia1996}. 

\begin{figure}[b]
\centerline{\includegraphics[width = 0.9\linewidth]{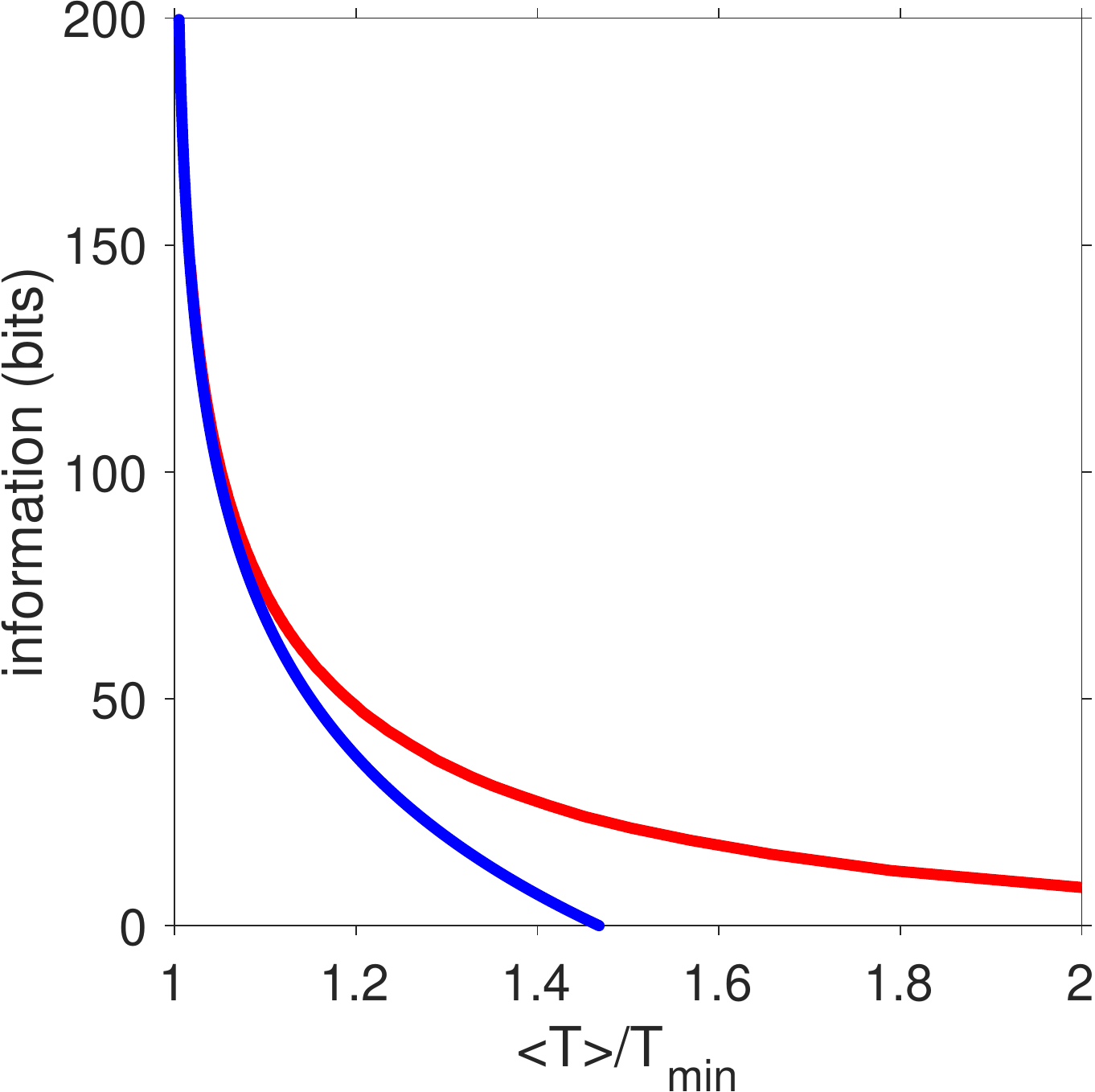}}
\caption{Information to specify tRNA abundances as a function of mean protein synthesis time per codon, with $\{f_{\rm i}\}$ from Fig \ref{fi_fig}.  Blue line is the Gaussian approximation, from Eq (\ref{Igauss}), and red line is the numerical result described in the text.  \label{IvsT}}
\end{figure}

If we just substitute the observed $\{f_{\rm i}\}$ into Eq (\ref{Igauss}), we find that getting within $\sim 5$\% of the optimum would require $\sim 100$~bits of information.  But  Eq (\ref{Igauss}) is an approximate result, only as good as our Gaussian approximation.  The condition for validity of this approximation is $\sigma_{\rm i} \ll t_{\rm i}^*$, or
\begin{equation}
{{T_{\rm min}}\over{\langle T \rangle - T_{\rm min}}} \gg {1\over{f_{\rm i}^{1/2}}} \sum_{\rm j} f_{\rm j}^{1/2}
\label{gauss_cond}
\end{equation}
for all $\rm i$. This condition is violated at $\langle T \rangle / T_{\rm min} \sim 1.2$, suggesting that we need to do better if we want to have a fully quantitative picture.  Happily, the partition function is exactly a product of Bessel functions [Eq (\ref{bessel})], and there are efficient algorithms for evaluating these functions.  Thus we can map $\ln Z(\lambda , \mu)$ in the $(\lambda, \mu)$ plane, evaluate derivatives by finite differences, impose the constraint in Eq (\ref{meant_con}), and then plot the information vs $\langle T \rangle$, parametrically in $\lambda$.    The results are shown in Fig \ref{IvsT}, including the Gaussian approximation for comparison.

We see that bringing the system within $\sim 10$\% of the optimal synthesis rate requires $\sim 50$~bits of information, or roughly one bit for each codon.  Even getting within a factor of two of the optimum requires $\sim 10$~bits.   One might worry that this is an overestimate, since we have assumed that each codon has its own complementary tRNA.  If we collapse down to 38 tRNA species \cite{collapse}, however, for $\langle T \rangle \rightarrow T_{\rm min}$ we find only a $10-15$\% reduction in the required information.    

It has been known for some time that the synthesis of fully functional tRNAs, charged with amino acids, involves many steps, all of which are subject to regulation \cite{altman_75}, and hence there are many paths along which the required information could be transmitted.   Each of these pathways will have a limited information capacity, and in the case of transcriptional regulation we know that this capacity is $\sim 1-2\,{\rm bits}$ \cite{tkacik+al_08a,tkacik+al_08b}, with precise numbers depending on the concentrations of transcription factors and the absolute copy numbers of the transcripts.  It is possible to imagine piecing together $\sim 10-50\,{\rm bits}$ of information along several pathways, but if real bacteria come close to their maximum translation rates then this may be possible only because they also come close to the information capacity of the relevant regulatory pathways.

\begin{acknowledgements}
We thank CG Callan, D DeMartino, and A Mayer for helpful discussions.
This work was supported in part by the US National Science Foundation, through the Center for the Physics of Biological Function (PHY--1734030), the Center for the Science of Information (CCF--0939370), and Grant PHY--1607612.
\end{acknowledgements}


\begin{thebibliography}{}
%
\bibitem{bialek_12}
W Bialek, {\em Biophysics: Searching for Principles} (Princeton University Press, Princeton NJ, 2012).
%
\bibitem{xia_88}
X Xia,   How optimized is the translational machinery in \textit{E~coli}, \textit{S~typhimurium}, and \textit{S~cerevisiae}? {\em Genetics} {\bf 149,} 37--44 (1988).
%
\bibitem{solomovici+al_97}
J  Solomovici, T Lesnik, and C Reiss, Does {\em Escherichia coli} optimize the economics of the translation process? {\em J Theor Biol} {\bf 185,} 511--521 (1997).
%
\bibitem{flux_opt}
RU Ibarra, JS Edwards, and BO Palsson, {\em Escherichia coli} k-12 undergoes
adaptive evolution to achieve in silico predicted optimal growth. {\em Nature} {\bf 420,}
186--189 (2002).
J Orth, I Thiele, and BO Palsson,  What is flux balance analysis? {\em Nat Biotechnol} {\bf 28,} 245--248 (2010).
%
\bibitem{demartino+al_16}
D De Martino, F Capuani, and A De Martino,  Growth against entropy in bacterial metabolism: the phenotypic trade-off behind empirical growth rate distributions in {\em E coli}. {\em Phys Biol} {\bf 13,} 036005 (2016).
D De Martino, F Capuani, and A De Martino, Quantifying the entropic cost of cellular growth control. {\em Phys Rev E} {\bf 96,} 010401 (2017).
D De Martino, Maximum entropy modeling of metabolic networks by constraining growth-rate moments predicts coexistence of phenotypes. {\em Phys Rev E} {\bf  96,} 060401 (2017).
%
\bibitem{neurons_maxent}
E Schneidman, MJ Berry II, R Segev, and W Bialek, Weak pairwise correlations imply strongly correlated network states in a neural population. {\em Nature} {\bf 440,} 1007--1012 (2006).
G Tka\v{c}ik, O Marre, D Amodei, E Schneidman, W Bialek, and MJ Berry II,  Searching for collective behavior in a large network of sensory neurons.  {\em PLoS Comput Biol} {\bf 10,} e1003408 (2014).
L Meshulam, JL Gauthier, CD Brody, DW Tank, and W Bialek,   Collective behavior of place and non-place neurons in the hippocampal network.   {\em Neuron} {\bf 96,} 1178--1191 (2017).
%
\bibitem{birds_maxent}
W Bialek, A Cavagna, I Giardina, T Mora, E Silvestri, M Viale, and A Walczak,  Statistical mechanics for natural flocks of birds.  {\em Proc Natl Acad Sci (USA)} {\bf 109,} 4786--4791 (2012).
W Bialek, A Cavagna, I Giardina, T Mora, O Pohl, E Silvestri, M Viale, and  AM Walczak, Social interactions dominate speed control in poising natural flocks near criticality.   {\em Proc Natl Acad Sci (USA)} {\bf 111,} 7212--7217 (2014).
%
\bibitem{proteins_maxent}
M Weigt, RA White, H Szurmant, JA Hoch, and T Hwa, Identification of direct residue contacts in protein--protein interaction by message passing. {\em Proc Natl Acad Sci (USA)} {\bf 106,} 67--72 (2009). 
DS Marks, LJ Colwell, R Sheridan, TA Hopf, A Pagnani, R Zecchina, and C Sander, Protein 3D structure computed from evolutionary sequence variation. {\em PLoS One} {\bf 6,} e28766
(2011). 
%
\bibitem{demartino+al_18}
D De Martino, AMC Andersson, T Bergmiller, CC Guet, and G Tka\v{c}ik, Statistical mechanics for metabolic networks during steady state growth. {\em Nature Commun} {\bf 9,} 2988 (2018).
%
\bibitem{jaynes_57}
ET Jaynes, Information theory and statistical mechanics. {\em Phys Rev} {\bf 106,} 620--630 (1957).
%
\bibitem{EhrenbergKurland1984}
M Ehrenberg and CG Kurland, Costs of accuracy determined by a maximal growth rate constraint. {\em Q Rev Biophys} {\bf 17,} 45--82 (1984).
%
\bibitem{limiting}
See  Ref \cite{solomovici+al_97} for a review.
%
\bibitem{bessel_ref}
IS Gradshteyn and IM Ryzhik, {\em Table of Integrals, Series, and Products}. Seventh edition, edited by A Jeffrey and D Zwillinger (Academic Press, New York,  2007).  See Eq (3.324.1). 
%
\bibitem{caglar+al_17}
MU Caglar, JR  Houser, CS Barnhart, DR Boutz, SM Carroll, A Dasgupta, WF  Lenoir, BL Smith, V Sridhara, DK Sydykova, DV Wood, CJ Marx, EM  Marcotte, JE Barrick, and CO Wilke, The {\em E coli} molecular phenotype under different growth conditions. {\em Sci Rep} {\bf 7,} 45303 (2017).
%
\bibitem{Sharp1993}
PM Sharp, M Stenico, JF Peden, and AT Lloyd, Codon usage: mutational bias, translational selection, or both? {\em Biochem Soc Trans} {\bf 21,} 835--841 (1993). 
%
\bibitem{Xia1996}
X Xia, Maximizing transcription efficiency causes codon usage bias. {\em Genetics} {\bf 144,} 1309--1320 (1996).
%
\bibitem{collapse}
If a single tRNA species is used for at most two codons, then the analysis above carries through with an effective $f_{\rm i}$ that is the sum of the two original values.  We carry out this projection from 61 codons to 38 tRNA species following Table 2 in H Dong, L Nilsson, and CG Kurland, {\em J Mol Biol} {\bf 260,} 649--663 (1996).  Small ambiguities in the projection are resolved as follows: the three Met tRNAs are treated as one, similarly for Thr1 and Thr3, Tyr1 and Tyr 2, and Val2A and Val2B.
%
\bibitem{altman_75}
S Altman, Biosynthesis of transfer RNA in {\em Escherichia coli}. {\em Cell} {\bf 4,} 21--29 (1975).
%
\bibitem{tkacik+al_08a}
G Tka\v{c}ik, CG Callan Jr, and W Bialek,  Information flow and optimization in transcriptional regulation.  {\em Proc Natl Acad Sci (USA)} {\bf 105,} 12265--12270 (2008)
%
\bibitem{tkacik+al_08b}
G Tka\v{c}ik, CG Callan Jr, and W Bialek,  Information capacity of genetic regulatory elements. {\em Phys Rev E} {\bf 78,} 011910 (2008).
%
\end{thebibliography}
\end{document}